# A Risk Assessment of a Pretrial Risk Assessment Tool

Tussles, Mitigation Strategies, and Inherent Limits

---

**Marc Faddoul[1], Henriette Ruhrmann and Joyce Lee**

School of Information, UC Berkeley

## ABSTRACT

We perform a risk assessment of the Public Safety Assessment (PSA), a software used in San Francisco and other jurisdictions to assist judges in deciding whether defendants need to be detained before their trial. With a mixed methods approach including stakeholder interviews and the use of theoretical frameworks, we lay out the values at play as pretrial justice is automated. After identifying value implications of delegating decision making to technology, we articulate benefits and limitations of the PSA solution, as well as suggest mitigation strategies. We then draft the Handoff Tree, a novel algorithmic approach to pretrial justice that accommodates some of the inherent limitations of risk assessment tools by design. The model pairs every prediction with an associated error rate, and hands off the decision to the judge if the uncertainty is too high. By explicitly stating error rate, the Handoff Tree aims both to limit the impact of predictive disparity between race and gender, and to prompt judges to be more critical of retention recommendations, given the high rate of false positives they often entail.

[1] Corresponding author at marc.faddoul@berkeley.edu

# TABLE OF CONTENTS

## BACKGROUND

The U.S. criminal justice system relies heavily on incarceration, accounting for under 5% of the total world population but over 20% of the total world prison population. The U.S. prison population rate is substantially higher than in many authoritarian countries, for example, twice as high as Iran's and five times higher than China's prison population rate[2]. Pretrial detention is a major factor driving mass incarceration, as 21.6% of the U.S. prison population are pretrial detainees[3], contradicting statements from the Supreme Court like "In our society, liberty is the norm, and detention prior to trial or without trial is the carefully limited exception."[4] In California, 63% of prisoners in jails have not been convicted or pled guilty, which raises the pretrial detention rate far above that of the rest of the U.S. Moreover, close to a third of felony defendants arrested are jailed but never found guilty of a crime[5].

Addressing the calamitous consequences of unjust pretrial detention is one of few policy issues with bipartisan support. From 2014 to 2015, San Francisco County spent over $3.2 million on jailing defendants who were never charged or whose charges were dropped or dismissed[6]; however, California's above-average pretrial detention rates have not achieved their intended results, as they have been associated with neither fewer defendants not appearing in court nor fewer rearrests[7]. Negative long-term impacts on an individual's professional and personal life in the criminal justice process are among the many consequences that are worse for defendants held in pretrial detention[8]. Black defendants are also disproportionately affected as the judicial system propagates racial disparities. Research has found less favorable treatment of black and latino defendants in comparison to white defendants by legal actors throughout the criminal justice process persists, even after controlling for legally relevant factors[9].

Many jurisdictions have thus taken action to move toward risk assessment tools to reduce the pretrial detainee population, address judicial bias in detention decisions, or remedy inequities due to monetary release conditions in the cash bail system. Risk assessment tools predict the statistical likelihood of uncertain future events, specifically the likelihood that a defendant will appear in court and not pose a danger to their community, based on information about the defendant. Risk assessment algorithms were intended to make access to pretrial release conditional on predicted risk in an objective process not susceptible to human bias rather than wealth, which was intended to benefit low income and minority defendants who were systematically disadvantaged in the prevailing criminal justice process. Such information about the defendant may include both invariable features and dynamic factors that vary over time[10]. At present, over 60 jurisdictions encompassing 25% of the U.S. population employ pretrial risk assessment tools[11].

---

[2] Roy Walmsley, "World Prison Population List" (World Prison Brief and Institute for Criminal Policy Research, June 11, 2018), http://www.prisonstudies.org/sites/default/files/resources/downloads/wppl_12.pdf.

[3] World Prison Brief, 'United States of America' http://www.prisonstudies.org/country/united-states-america [accessed 25 November 2018].

[4] Salerno v. United States, 481 U.S. 739, 755 (1987)

[5] Alison Parker, Danielle Haas, and Human Rights Watch (Organization), eds., *"Not in It for Justice": How California's Pretrial Detention and Bail System Unfairly Punishes Poor People* (New York, N.Y.: Human Rights Watch, 2017), https://www.hrw.org/sites/default/files/report_pdf/usbail0417_web_0.pdf.

[6] Joint Technology Committee, "Using Technology to Improve Pretrial Release Decision-Making," JTC Resource Bulletin (Conference of State Court Administrators, National Association for Court Management, National Center for State Courts, February 17, 2016), https://www.ncsc.org/~/media/files/pdf/about%20us/committees/jtc/jtc%20resource%20bulletins/it%20in%20pretrial%203-25-2016%20final.ashx.

[7] Justin Goss, "Pretrial Risk and Cash Bail," *Public Policy Institute of California* (blog), February 20, 2018, https://www.ppic.org/blog/pretrial-risk-cash-bail/.

[8] Joint Technology Committee, "Using Technology to Improve Pretrial Release Decision-Making."

[9] Matthew DeMichele et al., "What Do Criminal Justice Professionals Think About Risk Assessment at Pretrial?," SSRN Scholarly Paper (Rochester, NY: Social Science Research Network, April 25, 2018), https://papers.ssrn.com/abstract=3168490.

Yet the use of risk assessment tools is highly controversial because of the risk in replicating biases present in the data used to train the algorithm. Evidence of inherent racial discrimination present in Equivant's COMPAS risk assessment tool created a foundational case for the field of machine bias[12]. Amid rising concerns regarding unintended societal consequences of algorithmic systems, the Laura and John Arnold Foundation (LJAF) developed the non-profit Public Safety Assessment (PSA). Used in San Francisco, the PSA was designed to conscientiously avoid machine bias by striving to use "evidence-based, neutral information" and not to rely on "factors such as race, ethnicity or geography"[13].

In August 2018, California took the lead in passing Senate Bill 10 (SB10), which abolishes the cash bail system and requires counties to employ a "validated"[14] risk assessment tool. This formalizes an existing trend: as of 2015, 42 of 46 California counties with pretrial services (91%) already report using such a pretrial risk assessment tool. Most common are the Ohio Risk Assessment System-Pretrial Assessment Tool (ORAS-PAT), used by 38% of counties, and the Virginia Pretrial Risk Assessment Instrument (VPRAI), used by 36% of counties. Seven counties (15%) use a variety of tools, including COMPAS and PSA. Lastly, four counties (9%) rely on locally-validated risk assessment tools. The remaining five counties offer pretrial services but do not yet use a risk assessment tool[15]. However, the Judicial Council of California, charged through SB10 with compiling a list of validated pretrial risk assessment tools, expressed concern that only the PSA appears to meet the validation requirements set out in SB10[16].

## RESEARCH OBJECTIVE

The objective of our study is to analyze the value implications of algorithmic risk assessment tools for pretrial detention, in particular, the PSA, given its projected relevance for pretrial decision making in the wake of SB10. Our analysis seeks to highlight the limits of the PSA in its current application and discuss possible mitigation strategies to allow for value-based pretrial decisions in California. In terms of scope we limit our analysis to the functional handoff in pretrial detention decisions from cash bail system pre-SB10 to a risk assessment-based system post-SB10 in California.

---

10 DeMichele et al., "What Do Criminal Justice Professionals Think About Risk Assessment at Pretrial?"

11 Brandon Buskey and Andrea Woods, "Making Sense of Pretrial Risk Assessments," *The Champion*, June 2018, https://www.nacdl.org/PretrialRiskAssessment/.

12 Julia Angwin et al., "Machine Bias," *ProPublica*, May 23, 2016, https://www.propublica.org/article/machine-bias-risk-assessments-in-criminal-sentencing.

13 Leila Walsh, "Public Safety Assessment: A Risk Tool That Promotes Safety, Equity, and Justice," *Laura and John Arnold Foundation* (blog), August 14, 2017, http://www.arnoldfoundation.org/public-safety-assessment-risk-tool-promotes-safety-equity-justice/.

14 "Validation" for the purpose of SB10 means identifying and mitigating any implicit bias.

15 Californians for Safety and Justice and Crime & Justice Institute, "Pretrial Progress: A Survey Of Pretrial Practices And Services In California," August 2015, 20.

16 Tani G Cantil-Sakauye, Martin Hoshino, and Cory T Jasperson, "Senate Bill 10 (Hertzberg), as Amended March 27, 2017 – Letter of Concern Addressed to Assembly Member Jones-Sawyer," 2017, http://www.courts.ca.gov/documents/ga-position-letter-assembly-sb10-hertzberg.pdf.

## METHODOLOGY

Our methodology combines theoretical and empirical approaches to addressing our research objective. In an iterative process, we reviewed existing literature, conducted five expert and stakeholder interviews, and applied two theoretical frameworks considering the value implications of technological change. While many academics have addressed the question whether pretrial risk assessment tools lead to more favorable outcomes for defendants, literature on the opinions of stakeholders within the criminal justice system is less accessible. Researchers investigating the PSA on behalf of the LJAF have surveyed criminal justice professionals – including pretrial services staff, judges, prosecutors, and public defenders – to address this gap[17].

Our **expert and stakeholder interviews** allow us to augment these insights with qualitative data and include more perspectives. In addition to discussing the practical reality of pretrial risk assessment with the San Francisco Deputy Public Defender[18] who provided a sample PSA court report (see Appendix 1), our team interviewed a UC Berkeley and Dartmouth professor of Computer Science researching algorithmic risk assessment[19], and the Criminal Justice and Drug Policy Director of the ACLU of California[20]. Furthermore, we were particularly interested in understanding how the individuals most affected by the technology, criminal defendants, view the functional handoff. For this reason we interviewed two previously convicted individuals and conducted a card-sorting activity to elicit which information would be most relevant to accurately predict whether a defendant appears in court or reoffends. The card-sorting activity firstly prompted interviewees to identify information types based on their personal experience, before presenting information types currently being used in pretrial decision making processes, either as a PSA factor or in other regions of the world (Appendix 2). Our approach mirrored the LJAF's survey research[21] and complements the study's findings with insights from underrepresented stakeholders.

Finally, we structured our findings and analysis with the help of **theoretical frameworks**, the handoff model developed by Mulligan and Nissenbaum[22] and the Ethical OS toolkit developed by the Institute for the Future and the Omidyar Network[23]. We chose to employ the handoff model because the observed trend towards algorithmic risk assessments is an example of the redistribution of functionality for which the handoff model can provide systematic insights with regard to the values in play before and after task delegation. Furthermore, the Ethical OS toolkit allowed us to assume a future-oriented perspective and draw on our critical insights to devise targeted mitigation strategies to address value conflicts in the future.

---

[17] DeMichele et al., “What Do Criminal Justice Professionals Think About Risk Assessment at Pretrial?”
[18] Chesa Boudin, Interview with the Deputy Public Defender of the San Francisco Public Defender’s Office, December 2, 2018.
[19] Hany Farid, Interview with Professor for Computer Science, Dartmouth College and UC Berkeley, November 2, 2018.
[20] Margaret Dooley-Sammuli, Interview with the Criminal Justice and Drug Policy Director at the ACLU of California, November 16, 2018.
[21] DeMichele et al., “What Do Criminal Justice Professionals Think About Risk Assessment at Pretrial?”
[22] Mulligan and Nissenbaum, “Reasoning about the Values Implications of Reconfiguring Sociotechnical Entanglements: The Handoff Model.”
[23] Institute for the Future and Omidyar Network, “Ethical OS - A Guide to Anticipating the Future Impact of Today’s Technology,” 2018, https://ethicalos.org/wp-content/uploads/2018/08/Ethical-OS-Toolkit-2.pdf.

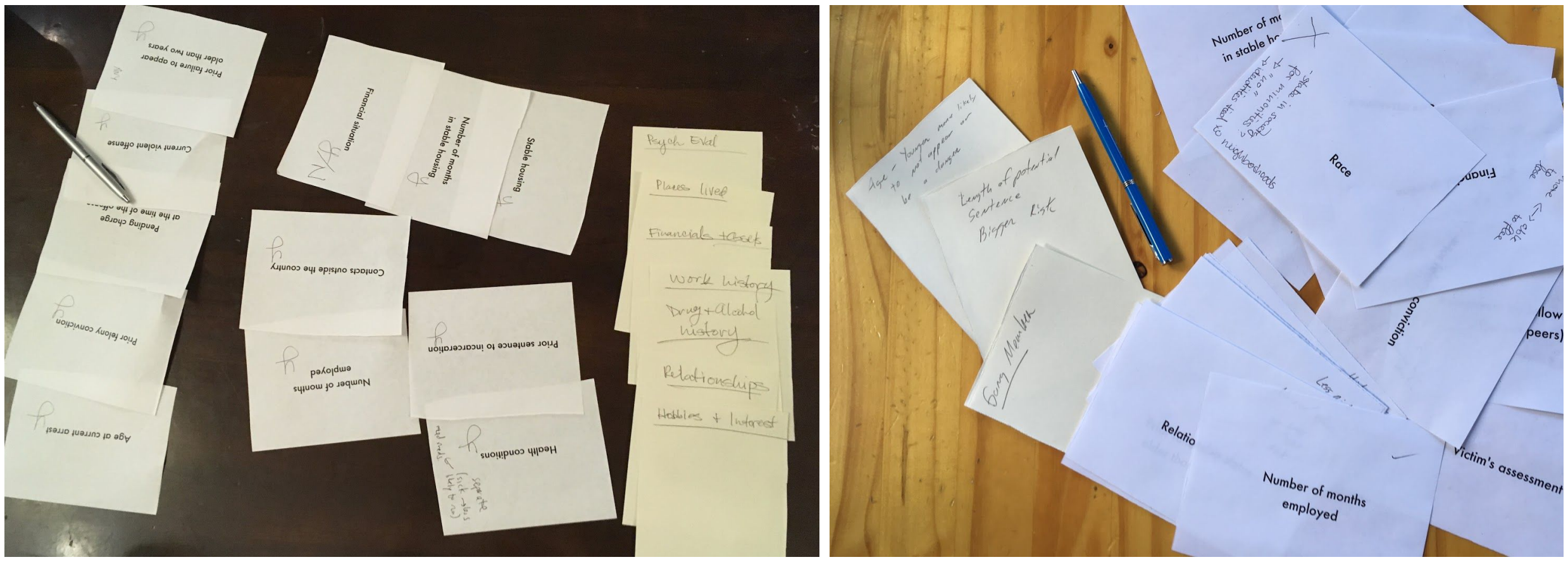

*Figure 1: Results of card-sorting activities analyzing information sources.*

# ANALYSIS

## Handoff Model

A "handoff" is a situation where a task or function is "shifted out" or delegated from one actor to another. Mulligan and Nissenbaum's framework contributes a cross-disciplinary analytical tool to map the values implications of a handoff. In our analysis, we apply the handoff model to two distinct handoff configurations (HC) in California pretrial decision-making before and after the enactment of SB10 (Table 1).

*Table 1: Handoff Configurations before and after SB10*

| **HC-A: Before SB10** | **HC-B: After SB10** |
|---|---|
| The judge decides whether to detain or release a defendant with or without conditions before the trial without algorithmic assistance. | The judge has the evaluation of the PSA available when deciding pretrial detention or release. |

The **system** for the purpose of our analysis is the judicial pretrial decision-making process at the level of the court, represented by the judge, and, in the case of HC-B, the risk assessment algorithm. The **trigger** for jurisdictions such as California to move from a non-technological pretrial decision-making process (HC-A) to algorithmic assistance (HC-B) may encompass two sets of factors: the availability of risk assessment algorithms and growing public frustration with extensive pretrial detention and inequitable access to pretrial release for financially insecure and minority defendants. The development of the COMPAS tool started in 1998 and was used outside a pilot project for the first time in 2010[24], while the LJAF began developing the PSA in 2013 and the tool has been in use in some jurisdictions as early as 2014[25].

## I. Goal

The system's **goal** under both handoff configurations is to ensure the defendant appears in court for their trial and does not pose a danger to their community for the duration of the criminal justice process. In achieving this dual goal, the system must satisfy the constraint of limiting unnecessary incarceration.

## II. Purpose

The **purpose** of pursuing this goal, ie. why the system seeks to achieve the goal, is to maintain public order and safety. Ensuring that a defendant appears in court is necessary for the defendant to receive fair treatment under the law, which includes a fair trial and if the defendant is convicted of a crime an appropriate sentence to maintain public order. Moreover, if an apprehended defendant poses a danger to the public, for example, because they threatened a witness in their case, it may be necessary to detain the defendant in the interest of public safety. The purpose of the constraint related to minimizing unwarranted incarceration is – at an individual level – to respect the defendant's presumption of innocence and limit restricting their liberty and – at a societal level – to avoid the human and financial cost of extensive incarceration.

## III. Function

In pretrial detention decisions, the **function** of the system is to assess all available information about the defendant and their case, to evaluate their risk of failing to appear in court and committing a crime that poses danger to their community. Based on predicted risk, the system must reach a decision whether to detain or release the defendant, with or without additional monetary or non-monetary conditions, for example, a GPS-monitoring device.

## IV. How

Both HCs discussed differ in **how** the system performs the function of assessing all available to predict a defendant's risk profile and reach a pretrial release or detention decision. The introduction of a risk assessment tool explicitly hands-off one aspect of the judge's decision to the algorithm: a numerical risk assessment.

## HC-A: Values Implications

The values inherent in HC-A reflect a high degree of agency both for the judge and the defendant. The judge enjoys full discretion and **autonomy** over the elements of the decision-making process and which part of the actual mental decision-making to publicly disclose to the defendant. Therefore, the judge is only constrained by the letter of the law and a possible legal challenge by the defendant. The fact that a judge can and will take into account qualitative information about the defendant's presentation illustrates the societal consensus that such information adds value to the decision. Bail schedules promote **efficiency** in the decision-making process and create a baseline for judges across courtrooms that anchors their decision on the conditions of release even as they retain full autonomy over the decision. To the degree that judges as human actors psychologically respond to anchoring, bail schedules contribute to harmonizing release conditions, which promotes **fairness** across courtrooms. However, a uniform bail schedule based on the charged offense creates a disadvantage for lower-income defendants because a nominally uniform bail amount impacts defendants charged with the same offense differently based on their income.

[24] Angwin et al., "Machine Bias."
[25] Laura and John Arnold Foundation, "About the PSA," accessed November 26, 2018, https://psapretrial.org/.

*Table 2: Analysis of Actors in Handoff Configurations A and B*

| **HC-A: Before SB10** | **HC-B: After SB10** |
|---|---|
| - The **human actor** (judge) has information on the defendant at an individual level and about similar defendants, cases, and their outcomes at an aggregate level, based on their subjective professional experience. The judge can take into account quantitative and qualitative information[26].<br>- The **human actor** (judge) defines which information is relevant and how it should be weighed. The judge may choose an intuitive or systematic approach (**mode**), for example, by applying a mental risk model. The judge is not required to distinguish between the independent goals of ensuring that a defendant appears in court and reducing the risk from future criminal activity to the public.<br>- The **human actor** (judge) arrives at a two-part decision: deciding whether to detain or release and whether to attach monetary (bail) or non-monetary (eg. electronic monitoring or abstinence from alcohol) conditions to a release, which the judge justifies to the defendant and their legal council. | - The **technical actor** (PSA) requires information deemed relevant to be entered into the system.<br>- **Human actors** (court staff) manually enter[27] information on nine distinct factors (**mode**) in the form of binary yes/no (eg. does the defendant have a prior conviction) or numeric inputs (eg. the defendant's age at current arrest)[28].<br>- The **technical actor** (PSA) calculates three risks scores[29] to predict whether a defendant will commit new criminal activity (NCA), commit new violent criminal activity (NVCA), or fail to appear in court (FTA)[30]. The PSA is explicitly oriented towards either goal because the algorithm is trained on appearance data or on recidivism data[31]. Each of the raw risk scores is converted to a six-point scale for the NCA and FTA scores and a flag, which it displays to the judge.<br>- The **human actor** (judge) has the same information available as under HC-A, the scores, and interpretation guidelines to make and justify the decision |

[26] The presence of the defendant's family, friends, or caseworker is considered "extremely" or "very" important in the decision to release or detain the defendant pretrial by 18% of judges. DeMichele et al. (2018)

[27] Manual data entry by a human actor is subject to intentional or unintentional human error and has in the past led inaccurate risk evaluations, for example, in the prominent case of a murder committed in San Francisco by a defendant assessed to be low risk who was released on bail. Lagos (2017)

[28] Interview with Margaret Dooley-Sammuli, the Criminal Justice and Drug Policy Director at the ACLU of California.

[29] Each of the three scores is based on a combination of four to seven of nine factors: age at current arrest, current violent offense, pending charges, prior misdemeanor conviction, prior felony conviction, prior violent conviction, prior failure to appear in past two years, prior failure to appear older than two years, and prior sentence to incarceration.

[30] Laura and John Arnold Foundation (2016)

[31] The LJAF discloses whether each factor impacts the risk positively or negatively and that "each of these factors is weighted—or, assigned points—according to the strength of the relationship between the factor and the specific pretrial outcome." The "strength of the relationship" between a given factor and one of the three risk outcomes was extrapolated from a dataset of observed outcomes in 1.5 million cases from 300 U.S. jurisdictions. The weights applied to each factor based on the model provided are numeric and range from 0 to 4 but mostly are binary pairs of 0 and 1 for yes/no-type factors. The process of calculating the three risk scores based on a prescribed formula by combining and weighting the information entered for each of the nine factors is deterministic in that the risk score output is fully determined by the data inputs and a set of static weights.

Moreover, the judges' autonomy in the decision-making process negatively impacts minority defendants, especially if their minority group is statistically more strongly associated with crime. In theory, the defendant is nevertheless provided with the judge's justification, which creates **transparency** and gives the defendant the opportunity to clarify misunderstandings and to challenge the judge's decision. Such transparency and space for communication between the judge and the defendant makes the judicial process **accessible** for the defendant and allows them to understand the process and exercise their rights. Furthermore, the decision-making process is also accessible to an appellate court and can be subjected to judicial **review** if there are concerns with regard to the quality of the decision. Finally, the judge is fully **accountable** for every aspect of the decision-making process, both de jure and in public perception.

## HC-B: A Shift in Values at Play

HC-B implies a shift in the values in play. The judge no longer has **autonomy** over which information enters the decision-making process because the nine risk factors that are evaluated in the PSA will factor into the decision. In the LJAF's survey, 80% of judges reported that the PSA "always" or "often" informs their release decision[32]. While it is possible for the judge to depart from the risk assessment, some jurisdictions have intentionally made such departures more procedurally costly, for example, by requiring the judge to state their reasons on the record[33]. SB10 in earlier versions while amended by the Assembly included the provision that "if a judge or magistrate's release decision is not consistent with the pretrial services program's risk assessment and recommendations on conditions of release, the judge or magistrate shall include in its order for release a statement of the reasons" (SB10, Section 17, §1275(c)(2017))[34].

Notably, this provision is not part of the final version of the bill as approved by Governor Brown: requiring a risk assessment tool as part of the decision-making process demonstrates societal values in prioritizing quantifiable quantitative information based on aggregated statistical data over qualitative information from the judge's professional experience. It is therefore not surprising that among legal actors interacting with the risk assessment tool, judges (33%) are most likely to consider the relative loss of judicial discretion a weakness of the system although both prosecutors (29%) and public defenders (25%) shared their concern[35]. Moreover, the process values **efficiency** as the input data is easy to collect and does not require interviewing the defendant. Risk assessment scores can then be calculated within split seconds, and in combination with the policymakers' guidance on the interpretation of risk scores provide a clear decision recommendation to the judge.

---

[32] DeMichele et al., "What Do Criminal Justice Professionals Think About Risk Assessment at Pretrial?"

[33] Harvard Law Review Note, "Bail Reform and Risk Assessment: The Cautionary Tale of Federal Sentencing," *Harvard Law Review* 131 (2018 2017): 1125.

[34] California State Legislature, "Senate Bill No. 10 (SB-10) Pretrial Release or Detention: Pretrial Services," Pub. L. No. 10 (2018), https://leginfo.legislature.ca.gov/faces/billTextClient.xhtml?bill_id=201720180SB10.

However, it is doubtful that HC-B can achieve its objective to promote **fairness** over the non-technological approach in HC-A. Even if race or income are not included in the PSA input parameters, 'prior misdemeanor conviction' or 'prior failure to appear' reflect societal and judicial biases against certain groups. The predictions of the PSA are therefore directly influenced by historical and present biases, just like judges can be. Rather than enhancing **transparency**, risk assessment tools often add obscurity to the decision-making process. Even though the LJAF stands out in its commitment to disclosing the risk factors and weights that the algorithm is based on, they do not disclose detailed information on the base dataset or the necessary subjective design choices in selecting seemingly simplistic integer weights to make the PSA easily describable. For example, an obvious case of a subjective design choice is the threshold effect created by the ''under 23 years old' penalty. For a defendant arrested one day before their 23rd birthday the NCA risk score is two points (out of a total of seven points) higher than if the same defendant had been arrested three days later.

This inconsistency is a result of a conscious choice to situate the PSA on a continuum between **interpretability** and **accuracy**. In other words, by thriving to fulfill public-facing values such as transparency and **accessibility**, the accuracy of the model was sacrificed. Nevertheless, the lack of transparent information on the model generation makes it more difficult to challenge a judge's decision following a risk assessment recommendation in judicial **review**. Finally, the public reaction to the case of the San Francisco murder on risk assessment recommended pretrial release illustrates that while the judge is still de jure **accountable** for the decision, public perception holds the technical actor responsible for adverse outcomes.

## Ethical OS Toolkit

The Ethical OS Toolkit features three tools to guide technologists toward "anticipating the future impact of today's technology (or: how to not regret the things you will build)."[36] We began our analysis by contemplating scenario six of the first tool, which happens to mirror the topic of this report: the adoption of "predictive justice" tools to determine prison sentences (Table 2).

Following this, we applied "The Risk Mitigation Manual," examining questions posed of risk zone four (machine ethics and algorithmic biases). The last two questions were particularly resonant: *"How will you push back against a blind preference for automation (the assumption that AI-based systems and decisions are correct and don't need to be verified or audited)?"* and "Are

[35] DeMichele et al., "What Do Criminal Justice Professionals Think About Risk Assessment at Pretrial?"
[36] Institute for the Future and Omidyar Network, "Ethical OS - A Guide to Anticipating the Future Impact of Today's Technology."

*your algorithms transparent to the people impacted by them? Is there any recourse for people who feel they have been incorrectly or unfairly assessed?*" Our interview with a representative from the San Francisco Public Defender's Office confirmed the former concern, highlighting how judges follow "do not release" PSA recommendations in over 90% percent of cases. Furthermore, he suggested that within the jurisdiction of San Francisco, legal counsel of current defendants have access to PSA recommendations; however, the degree to which they are explained to defendants likely varies on a case-by-case basis.

Table 2: Contemplating Scenario 6 of the First Ethical OS Tool

| | |
|---|---|
| **What is your greatest worry in this scenario?** | Systemic injustice is perpetuated, social change is impossible, and norms are inverted, shifting toward "guilty until proven innocent" |
| **How might different users be affected differently by this future?** | Certain individuals and communities that are overpoliced are more negatively affected, either inadvertently or intentionally |
| **What actions would you take to safeguard privacy, truth, democracy, mental health, civic discourse, equality of opportunity, economic stability, or public safety?** | Prevent automation of decision making processes by requiring justification of decisions made, make records of such decisions and justifications publicly available |
| **What could we be doing now to get ready for this risky future?** | Shape regulatory mechanisms (technology, law, market forces, social norms)[37] accordingly: require regular algorithmic audits, implement support better training for judges, incentivize transparency to align with business objectives, promote social accountability and opportunities for contestability |

Examining the final future-proofing tool, we dwelled in the possibility of strategy three (Ethical Bounty Hunters); however, it would require transparency of datasets and algorithmic inner workings to be viable. As a result, we turned to strategy two, A Hippocratic Oath for Data Workers, which features Data for Democracy's Global Data Ethics Pledge (GDEP).[38] We imagined the oath being taken by algorithm designers in courts of law, committing to the five principles of fairness, openness, reliability, trust, and social benefit, before presenting algorithms in regular audits for public scrutiny. Despite its forward-looking framing, the toolkit's emphasis on avoiding regrettable mistakes is noteworthy; indeed, after examining each tool within the context of algorithmic risk assessment, we found that the framework aims to prevent repetition of past mistakes more so than foster truly out-of-the-box speculation.

[37] Lawrence Lessig, "Chapter 7 - What Things Regulate," in Code, Version 2.0 (New York: Basic Books, 2006).
[38] Data for Democracy, "Global Data Ethics Pledge (GDEP)." 2017. https://github.com/Data4Democracy/ethics-resources

# DISCUSSION

Building up on the insights gathered from existing literature, analytical frameworks, and stakeholder interviews, the following section highlights both benefits and limitations to the PSA, as well as possible mitigations. We expand the discussion to consider a few broader considerations of algorithmic risk assessment, before proposing an alternative design, the Handoff Tree Model, to address identified constraints.

## PSA Benefits and Limitations

We acknowledge that the PSA solution was built in good faith with the intention to diminish unnecessary pretrial incarceration. Compared to other tools, it does indeed offer a few benefits. The PSA uses data that is easy to collect and is thus easy to implement, not requiring the defendant to be interviewed. This contributes to making it a cost-effective solution and avoids introducing personal biases of the interviewer in an interview with the defendant. Moreover, our interview with the San Francisco Public Defender's Office confirmed that more people are being released within San Francisco – both at pre-arraignment and early stage – as a result of using the PSA.

Unlike Northpointe's COMPAS tool, the PSA also breaks down the risk scores to make the weight of each variable explicit, and it separates three types of risks – failure to appear (FTA), new criminal activity (NCA), and new violent criminal activity (NVCA). Furthermore, the tool intends to rely on data that limits the impact of structural biases against minorities. Variables such as postal code, income, or juvenile convictions that are highly correlated to race are not included. Even if this does not make the tool immune to bias as suggested by LJAF, the PSA seems to be essentially calibrated between races and gender. Calibration means that for a given risk score, there is a similar chance of FTA, NCA, or NVCA if the defendant is released, independently of the race. Offered a dataset from Kentucky courts by the LJAF, researchers' assessments found that PSA is relatively well calibrated between white and black defendants, showing minimized predictive biases for two of the three risk scores.[39] Nonetheless, design choices made by the LJAF challenge several values, including **fairness**, **transparency**, **precision**, and **empirical accuracy**.

### Inherent Disparities Are Not Explicitly Addressed

Given that some racial groups have more prior charges, all subject matter experts interviewed expressed the problematic nature of using flawed historic data to create a model such as the PSA. Effective calibration means that for a given risk score, there is a similar chance of FTA (or NCA, NVCA) if the defendant is released, independently of the race. While meaning well, **calibration efforts do not remove predictive bias**: because the average number of defendant who fail to appear or recidivate (base rates) are different across races and genders, there is an inherent mathematical tradeoff between calibration and error-rate balance.[40] This is confirmed by the aforementioned analysis of the PSA done with the Kentucky dataset, which found *"the FTA scale to be significantly more predictive for white defendants (ROC = 0.655) than black defendants (ROC = 0.612)."*[41] ProPublica and others' critiques of the COMPAS tool also thus apply to the PSA: a well intended black defendant is more likely to be predicted as high risk than a equally well intended white defendant (essentially, a higher false positive rate for black people). The same phenomenon occurs between genders for the NVCA score, making the PSA harsher for women.

[39] The exception is for low-risk black defendants, whose failure to appear (FTA) rate tends to be underestimated.

Predictive disparity across race and gender is inherent and needs to be explicitly addressed; however, the LJAF fails to state how the trade-off between error-rate and calibration is handled. The stance for calibration was only inferred from the Kentucky dataset, which was picked by LJAF. Without a fairness proposition, there is no guarantee that the PSA will remain calibrated over an extensive dataset. Moreover, it is an intentional choice from the LJAF not to include race in the prediction, but some research argues that blinding the algorithm to race and other protected variables only detracts from fairness metrics.[42] On the contrary, **including race or gender allows to reach optimal precision both within groups and overall.**[43] These theoretical results concur with the PSA Kentucky analysis, which found that including the 'race' variable would add predictive utility.[44] Unfortunately, politicians are reluctant to accept distinctions by race or gender despite the statistical benefit. Treatment disparity is often illegal[45] and always a sensitive topic.

***Mitigations.*** Approaches to enhance fairness would thus be for the LJAF to state an explicit fairness proposition regarding racial and gender bias. This might include adjusting the PSA to include race and gender in jurisdictions where it is legal, and take position in favor of treatment disparity to increase predictive utility.

## The PSA's Underlying Statistical Model is Opaque

Without transparent judicial processes, the state cannot be held accountable for fairly enforcing the rule of law. Paradoxically, despite the fact that the LJAF supports "a deliberately transparent approach"[46], **neither the data nor the model used to infer the weights and features of the model are public.** Since the LJAF made the choice to only include input parameters that cannot be tampered with by the defendant or the prosecutor (see Appendix 2 for PSA factors), this proprietary opacity cannot be justified with an intention to protect the tool from being outsmarted by the actors who use it. As suggested by our conversation with the San Francisco Public Defender's Office, this deliberate information retention is likely a way for the LJAF to protect their solution from competitors and to limit exposure to criticism. It is rather concerning that market regulation principles such as trade secrets are preferred over accountability and transparency of the judicial system.

[40] Jon Kleinberg, Sendhil Mullainathan, and Manish Raghavan, "Inherent Trade-Offs in the Fair Determination of Risk Scores," *ArXiv:1609.05807 [Cs, Stat]*, September 19, 2016, http://arxiv.org/abs/1609.05807.
[41] Matthew DeMichele et al., "The Public Safety Assessment: A Re-Validation and Assessment of Predictive Utility and Differential Prediction by Race and Gender in Kentucky," *SSRN Electronic Journal*, 2018, https://doi.org/10.2139/ssrn.3168452.
[42] Jon Kleinberg et al., "Algorithmic Fairness," *AEA Papers and Proceedings* 108 (2018): 22–27, https://doi.org/10.1257/pandp.20181018.
[43] Zachary C. Lipton, Alexandra Chouldechova, and Julian McAuley, "Does Mitigating ML's Impact Disparity Require Treatment Disparity?," *ArXiv:1711.07076 [Cs, Stat]*, November 19, 2017, http://arxiv.org/abs/1711.07076.
[44] DeMichele et al., "The Public Safety Assessment."
[45] The SB-10 bill is in contradiction with statistical evidence, as it requires risk assessment tools to be 'equally accurate' across gender or race, but prohibits the use of these variables. Though the bill can still be changed, and these constraints are state dependant. The Wisconsin supreme court has authorised gender-specific risk assessment tools in a 2016 ruling.
[46] "Pretrial Justice - Laura and John Arnold Foundation." http://www.arnoldfoundation.org/initiative/criminal-justice/pretrial-justice/. Accessed 9 Dec. 2018.

Another key challenge with the PSA's opacity is that there is no clarity for how recommendations are made. In PSA-generated court reports, for instance, judges are offered the input data, the resulting NCA and FTA scores, and a table to get the final recommendation (see Appendix 1). The connection from raw scores to recommendations is not transparent, which prevents judges and – often policymakers at large – from understanding the PSA, generating a misunderstanding of algorithmic capabilities. Opacity in the functional logic of the tool contributes to the "artificial intelligence" mythology that overstates the power of prediction tools. As our interview from the ACLU articulated, *"[Computerized] risk assessment sounds like a silver bullet, but there's a lot of education to do."*

***Mitigations.*** A key improvement to improve transparency would be to publish the underlying statistical model that was used to generate the weights associated with each input variable, as well as the methodology used to clean the data and generate the training labels. Anonymised subsets of the training data should also be ideally included to better enable accountability. Furthermore, providing a clear justification for how risks scores are converted into the recommendation matrix would making the model more understandable for all the actors involved.

## The PSA Model is Too Simplistic, Reducing Reality's Complexity

The PSA limits its model to a sum of binary variables to reduce complexity. It seems that the approach is aimed at increasing the interpretability of the decisions, but it results in an overly simplistic model. Firstly, this creates **arbitrary threshold effects**: for instance, a defendant suddenly loses two (out of 13 raw risk points) on the NCA score as they turn 23 years old. Secondly, summing up risk factors also implicitly assumes that they correspond to independent phenomenons and that there are **no cross-variable effects**. This is a debatable hypothesis, as there most likely are interactions between age and prior convictions. Thirdly, a fundamental aspect of empirical science is to include error rates and confidence intervals. The PSA includes **no considerations regarding the confidence of the results**, despite the fact that it cannot be identical for all defendants. For instance, one can intuitively notice that the precision of the model will be lower for defendants close to 23 years old because of the threshold effect described above. In sum, the over-simplification of the PSA formula leads to aberrations and potential blind spots without providing feedback on the precision.

***Mitigations.*** To avoid arbitrary penalties for defendants as a result of the model's lack of nuance, mitigation strategies may include smoothing empirically irrelevant threshold effects, providing confidence intervals, and accounting for interactions between key variables by explicitly including or controlling for them in the model.

### Use of "Red Flags" Countervails Empirical Evidence

The PSA has a feature called “step-two exclusions”: these are overrides that lead to systematic “do not release” recommendations. By default (as a a result of LJAF’s reasoning), a murder charge is one such “red flag,” but counties also have the capacity to implement their own step-two exclusions, as we learned by way of our interview with San Francisco Public Defender’s Office. These flags are often introduced based on human intuition and societal preference; however, **is not the point of risk assessment tools is to challenge human intuition with empirical evidence?** The tool should be made to pick up on overlooked phenomena: for instance, that in the case of a spousal homicide the murderer is often not a risk to the rest of society, or that many may defendants are actually innocent – a common problem when using charges as an input. Moreover, judges are well able to enforce a systematic exclusion by themselves without a built-in override through the PSA. There is no value in using a machine to perform a trivial flagging task based on charges. Charges are not currently included in the input. If specific charges need to be part of the formula, it should be done in a empirical way rather than hard-coding human judgements, which undermines the whole predictive entreprise.

***Mitigations.*** We question the nature of step two exclusions – we intended to speak with the LJAF about this decision, though they did not return for comment. Without further context, we suggest the removal of step-two exclusions altogether, or at least the removal of the ability to create more automatic overrides: our ideal alternative would be to provide an empirical risk score, and leave to the judge his agency and autonomy to override the risk score, if necessary. We elaborate further on this in our discussion of the Handoff Tree Model design.

## Broader Algorithmic Risk Assessment Considerations

### Prediction Accuracy in Pretrial

The PSA’s overall predictive utility is about 64%, a level that matches other algorithms like COMPAS[47]. This can sound decent at face value; however, some perspective on this figure may be helpful. A 64% precision is equivalent to saying that if we randomly pick one released defendant that appeared to court and one released defendant that failed to appear, the PSA algorithm would have attributed a higher score to the latter 64% of the time. This is just 14% better than flipping a coin. A recent experiment gives another interesting point of comparison. Participants with no judicial expertise were asked to guess whether a defendant would recidivate based on the same input variables that the COMPAS algorithm uses. They had a median accuracy of 64% where the COMPAS algorithm yields 65.2%.[48] Are risk assessment tools really introducing valuable insights?

---

[47] DeMichele et al., “The Public Safety Assessment.”

[48] Julia Dressel and Hany Farid, “The Accuracy, Fairness, and Limits of Predicting Recidivism,” *Science Advances* 4, no. 1 (January 2018): eaao5580, https://doi.org/10.1126/sciadv.aao5580.

Notably, the base rate for recidivism within that dataset was 46%, which means that the algorithm did much better than a 54% baseline one could reach by trivially guessing “no recidivism” all the time. For failure to appear, the base rate is 14.8% in the Kentucky dataset. **This means that the PSA algorithm would be vastly outperformed by an ”algorithm” that would recommend release all the time, which would achieve 85% accuracy.** The Kentucky dataset includes defendants who have been released, but for whom the PSA algorithm would have given a “retain” recommendation. The empirical offense rates for these high risk profiles are outlined in Table 3. In other words, if a defendant is classified as 'retain' for having a FTA score of 5, there is empirically a 75% chance that they will NOT fail to appear. **For new violent criminal activity, this false positive rate goes up to 97%**. Do the misbehaviors of the remaining 3-4% justify the mass incarceration of the rest of these defendants?

*Table 3: Empirical Offense Rates of High-Risk Profiles (According to the PSA)*[49]

| PSA Risk Score (out of 6) | 5 | 6 |
|---|---|---|
| Failure To Appear | 26% | 32% |
| New Criminal Activity | 20% | 26% |
| New Violent Criminal Activity | 3% | 4% |

We concede that this dataset is limited to Kentucky, and that we only have data on defendants for whom the counterfactual is available, which are defendants who have been released. Even if this may not be perfect, we trust LJAF for having provided a dataset that is roughly representative of reality. Therefore, even if the exact percentage may vary, the order of magnitude should be correct. The quality of the dataset provides the same credit to these base rate estimations than to the fact that the PSA is well calibrated regarding race. Even if the base rate were to be 6% instead of 3%, the table shows that for optimal precision, all defendants should be released all the time. Prediction tools, including PSA, legitimise the fact that for every 4 prisoners sent to jail for FTA risk of 5, only one would have defected and the other 3 are purging unnecessary pre-trial sentences.

## The Logistical Difficulties of Qualitative Information

The PSA’s emphasis on being efficient means that it prioritizes data that is easy to collect. However, during our card-sorting exercises with formerly incarcerated individuals, research participants and their lived experience identified primarily qualitative information sources as being useful to making risk assessments (see Figure 1). A few of these include psychiatric evaluation (via the Global Assessment of Functioning Scale), number and quality of relationships, as well as length of potential sentence – with longer sentences leading to riskier behavior, mirroring projections of prospect theory.[50] The only unprompted information source that mirrored the PSA risk factors was age, which one of the participants did identify. Many of the qualitative information would only be accessible by a face-to-face interview or other forms of investigation. However, not only does this require resources, but it may also introduce bias which the algorithmic risk assessment tools were intended to alleviate.

[49] Re-crunching of the data presented in DeMichele et al., “The Public Safety Assessment.”, Table 20, 21 and 22.
[50] Kahneman, Daniel, and Amos Tversky. “Prospect Theory: An Analysis of Decision under Risk.” *Econometrica*, vol. 47, no. 2, 1979, pp. 263–291. *JSTOR*, JSTOR, www.jstor.org/stable/1914185.

This tradeoff is particularly notable when attempting to understand whether a failure to appear in court is intentional or unintentional, as relevant information often cannot be efficiently gathered. *"There's a big difference between flight risk (e.g. someone running away to Mexico) vs. failure to appear risk (so, say if I'm homeless and I don't have a calendar or phone),"* one of our interview subjects explained. *"[Algorithmic] tools don't distinguish between those. If 'runaway to Mexico' risk is very hard to manage; that's a very different situation from someone who is 'high risk' of failure to appear because they had no one remind them, give them bus tokens, etc. The tool conflates these risks, some of which can be easily mitigated."* A number of external circumstances can affect whether or not someone who fails to appear in court, but understanding any defendant's intentions faces challenges of not only gathering such qualitative data, but also incorporating it the algorithm, as a question of efficiency and practicality.

## ALTERNATIVE DESIGN SUGGESTION: THE HANDOFF TREE MODEL

Given the aforementioned PSA limitations and broader considerations, we suggest an alternative algorithm design. It does not significantly increase the precision, as the low base rate and high uncertainty are inherent. However, the core idea is that **the algorithm only makes a prediction when the confidence is high enough**; when it is not, the decision is handed off to the judge. It is a process of **intelligent delegation,** in which the algorithm calls upon the critical judgement of the judge when it is most needed, preventing blind deference to algorithmic recommendation.

Benefits of such a design include diminishing the error-rate, as the algorithm will not make excessively noisy predictions. **Predictions are always bundled with an error rate**, to precise what is meant by 'high' or 'very low' risk and increase the interpretability of the tool. Moreover, by making the false negative rate apparent, a dimension of valence is added to the predictions which **addresses the predictive disparity across races and genders.** Furthermore, the tree approach offers greater **accountability** in that the combination of factors that led to a prediction are trackable and can be debated.

### Prediction Process of the Handoff Tree

A decision tree is an intuitive, yet powerful model. Unlike the PSA model, a decision tree makes no assumption of linearity and captures the interactions between the risk factors. Decision trees are widely used for their ability to integrate large heterogeneous datasets while minimizing the loss of relevant signal.

To derive a prediction, a series of questions is asked about the defendant. Let us consider a toy example with only two variables: we want to predict the risk of failure to appear based on the age and the number of prior failures to appear of a defendant (see Figure 2). The blue circles (Os) correspond to defendants who have been released and came to court, while the red crosses (Xs) are defendants who failed to appear. This graph represents the training set, or knowledge base of the algorithm. Running a decision tree algorithm will divide the cloud of Xs and Os into clusters – represented by the colored rectangles, each of which corresponds to a prediction label.

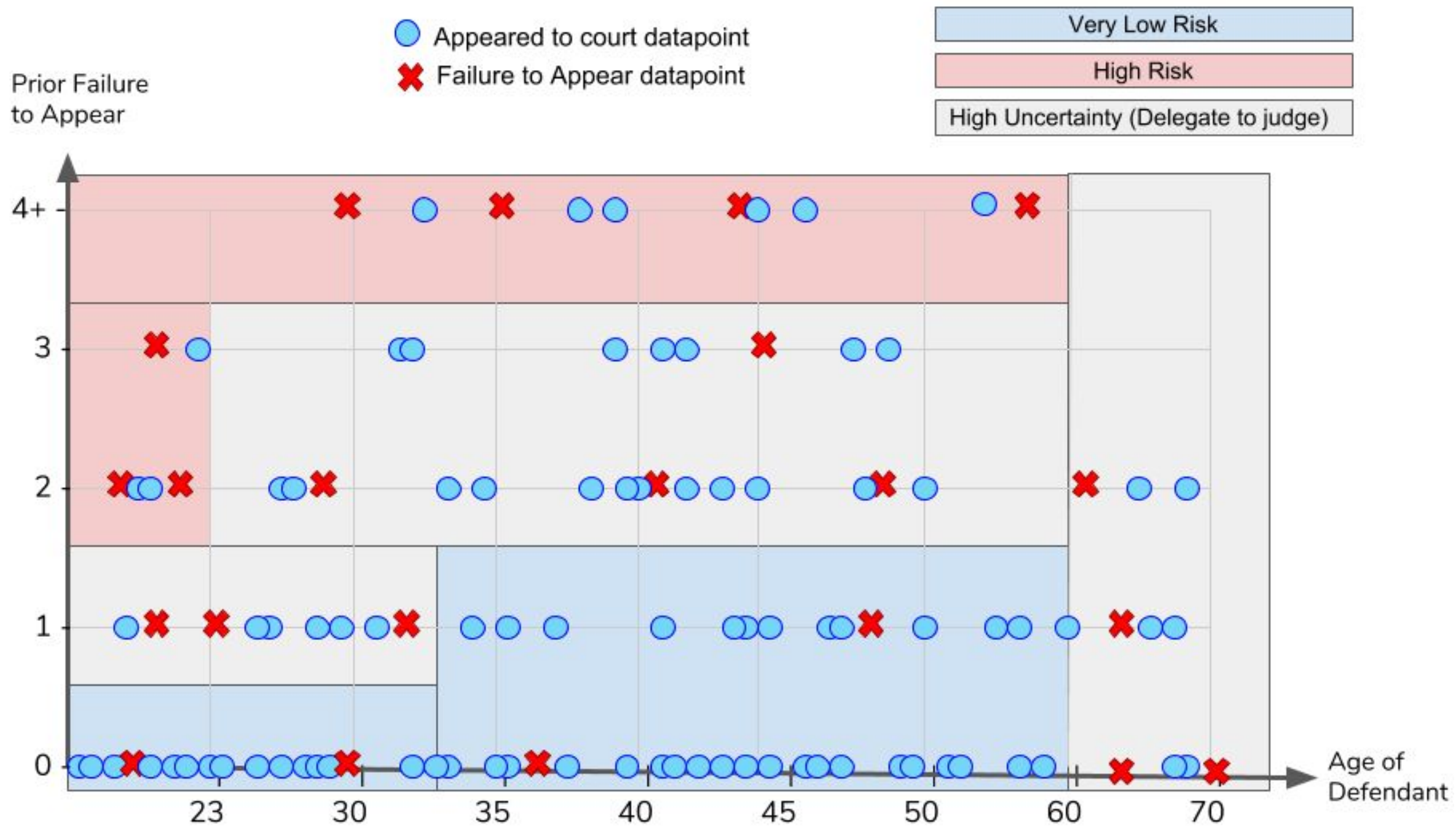

*Figure 2: Representation of the Handoff Tree on the training dataset.*

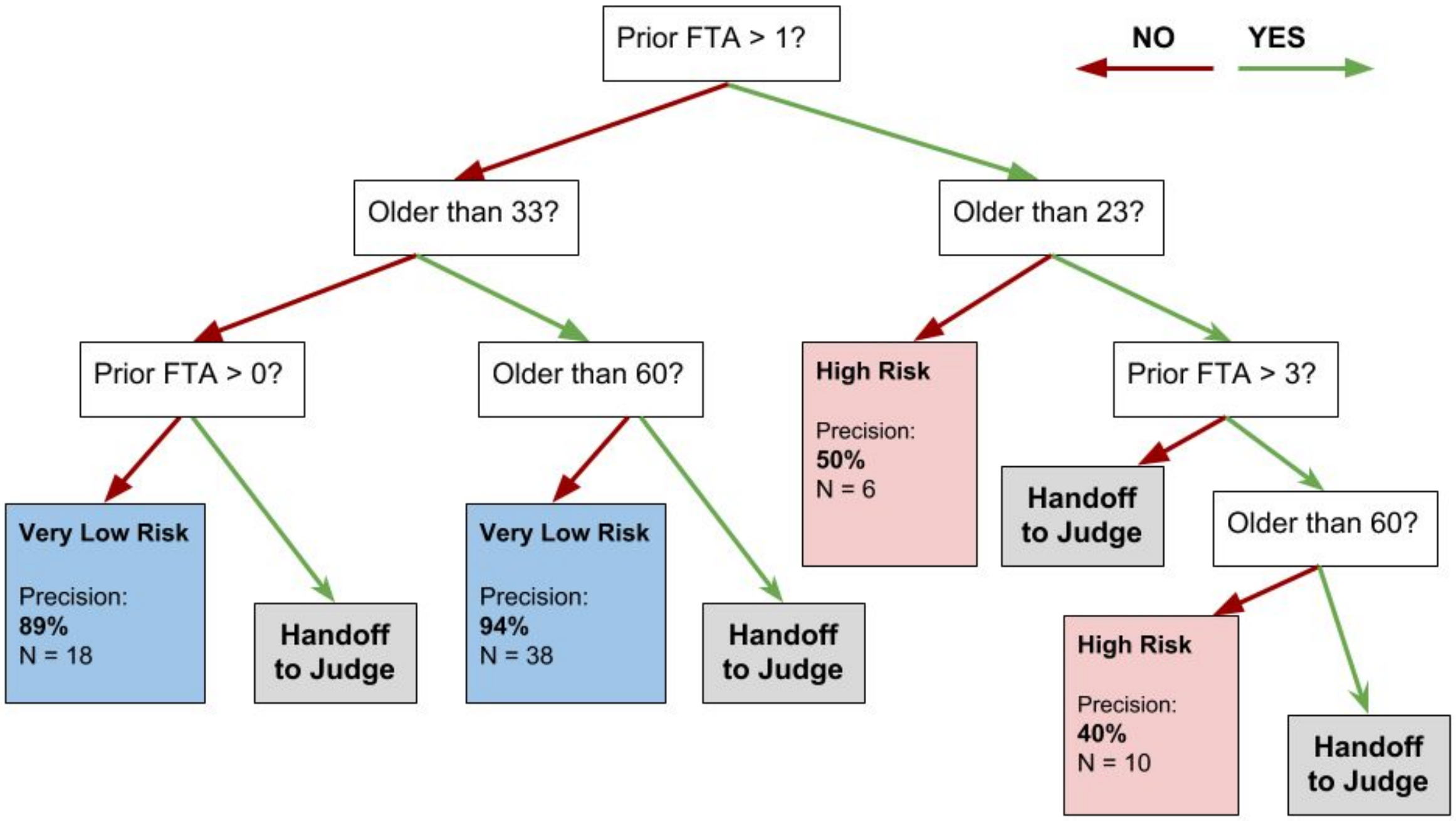

*Figure 3: Representation of the Handoff Tree decision process*

The algorithm follows an ordered list of questions, derived by the training dataset: for instance, Is the defendant older than 33, or older than 60? Do they have more than one failure to appear on their record, or more than three failures to appear? With each question, the profile of the defendant takes form, enabling the algorithm to curate a list of similar defendants from the training dataset, to whom it may compare the defendant in question. This list, or cluster, corresponds to the colored rectangles on Figure 3.

Some clusters are very well defined: for instance, almost all defendants between 33 and 60 who had never failed to appear before did show up to trial. In these regions of the graph, the algorithm can give a prediction with a strong confidence. On other parts of the graph, however, the training set shows much more uncertainty. For instance, for defendants under 33 with a single prior failure to appear, there is no strong majority in the training set. Instead of giving an average risk, **the algorithm hands off decision to the judge, who can leverage contextual elements to refine the prediction.** The judge is more properly suited to investigate, asking the defendant why they failed to appear the first time and assessing if the cause can be mitigated.

A key design consideration is that **the model only gives an empirically-based risk prediction, without any recommendations**, a contrast to the PSA's approach, which goes beyond risk assessment to also make recommendations. This is an improvement as it eliminates decision making on data that has only a low statistical confidence. To address the low base rate of FTA and NCA which causes high rates of false positives, we explicitly state error rates associated with each prediction. The handoff tree allows to report confidence easily, by simply taking the ratio of positive to negative points in a given cluster. For instance, in the "High Risk" cluster of defendants with 4+ prior failures to appear, the false positive rate is 60%. In the "Very Low Risk" cluster of individuals under age 33 with no prior failures to appear, the false negative rate is 13%. We hope that these rates will inform the judge's decisions, by systematically reflecting the fact that high risk predictions often correspond to an empirical risk far below 50%.

Lastly, **making the false negative rate apparent is a way to address the problem of discriminatory bias.** As described in the discussion on fairness and calibration, the core of the problem is that for the tool to be calibrated, black defendants have higher false positive rates than white defendants: to address this, the error rate will add a notion of valence to every prediction. **Black defendants might still be more often classified as "high risk" than their white peers, but the higher error rates will make the strength of these prediction weaker.** If judges integrate these rates in their decisions, this strategy could concretely mitigate the discriminatory biases relative to the impossibility of simultaneously accommodating fairness of treatment and fairness of impact.

## Values Implications in Designing the Handoff Tree

Code is law, or, as Winner puts it, "artifacts have politics"[51]. Indeed, the **technical choices made during the design of the model entail strong value implications.** The handoff tree model is no exception to this position: the choice of the training set and its features as well as the value of several hyper-parameter will reflect political or ethical judgements.

### Shaping the Training Set and its Features

First we would like to highlight that **only defendants who have been released can be used as training data.** For defendants who were put in jail, there is no counterfactual: we don't know what they would have done if they had been released. Therefore, they can't be used as training data. We assume that the PSA adopted this approach, but we did not find an explicit statement that confirms it. Another important design choice is whether race and gender should be integrated in the feature set. As stated mentioned above in the discussion about fairness and calibration, we believe that the predictive signal introduced by these highly relevant variables is valuable to reduce predictive disparity and increase accuracy. As much as it is legally possible, we argue that **it is beneficial to account for gender and race while clearly stating the fairness proposition that support the use of these protected variables.**

Another consideration regards the number of prediction tasks the model should include. The PSA has three, separating the risk of failure to appear, the risk of commiting a crime before the trial and the risk of committing a violent crime. This is a wise choice, but another important distinction is lacking, as suggested by several of our interviewees: **is it a risk of *intentional or unintentional* failure to appear?** Indeed, it seems that many defendants fail to appear because of transportation problems or conflicts with professional or personal constraints. Unlike for defendants who are trying to flee, the appropriate treatment for a high risk of unintentional failure to appear is not incarceration. Instead, one would want to find ways to have the defendant commit to coming to court or to accommodate material constraints. It seems that providing a bus ticket could sometimes save months in jail.

[51] Winner, Langdon. "Do Artifacts Have Politics?" Daedalus, Vol. 109, No. 1, Modern Technology: Problem or Opportunity? (Winter, 1980), pp. 121-136 The MIT Press on behalf of American Academy of Arts & Sciences.

The problem with that approach is that constituting a training set that would distinguish between intentional and unintentional FTA would require a significant amount of investigative work. Then, one would want to identify features that can be collected easily and that predict the FTA intentionality well enough. For instance, “accessibility of the court by public transportation” or “date of trial during work hours of the defendant” could be relevant criteria. Discrimination between intentional or unintentional FTA could be done both on the PSA model and on the handoff model. It is not low-hanging fruit, requiring a substantial amount of work both to design and implement the process. Nonetheless, we believe it is worth exploring as this change could substantially alleviate unnecessary incarceration.

## Setting Thresholds and Other Hyperparameters

The hand-off tree is a generic model, and there are several hyperparameters that will impact the behavior of the algorithm in practice. The first parameter is **the precision threshold below which the algorithm will delegate the decision to the judge.** Since the low base rate for FTA and crimes before trial make the classification asymmetrical, there are two primary thresholds to set: (1) the minimal false positive rate in order to classify a defendant as “high risk,” and (2) the minimal false negative rate in order to classify a defendant as “very low risk.” Choosing these values will impact how much jurisdiction the algorithm takes upon itself. Most importantly, it **will define the extent to which our society accepts to put well-intended defendants in jail in order to reduce FTA and pretrial crime rates.**

Another needed specification regards **how the clusters are defined during the training phase.** For instance, do we want to also consider defendants that are aged 28 when we are predicting the risk of a 22-year-old defendant? Or would it be better to restrict the inference to a cluster of defendants in the age range 18-26? The bigger the clusters are, the less treatment disparity there is between defendants. In general, it will make the predictions safer, but less precise. When the cluster size is reduced, the predictions become more relevant as we compare defendants that are more alike. The downside of this is that as the sample size diminishes, the model becomes more prone to doing undue extrapolations. Choosing an excessively small cluster size would correspond to what is called *overfitting* the model. Picking the optimal cluster size depends on the size and quality of the training-set. This optimisation is an intricate statistical problem involving trade-offs between:

- the **confidence** and **precision** of predictions
- the **computational complexity** of the model (especially relevant if using the later discussed forest extension of the handoff tree)
- **data confidentiality**, or the degree to which the training set can be inferred from the model

Formalizing how these technical constraints can be translated into their implications regarding judicial values would require further work. Furthermore, even after having set the cluster size and all other hyperparameters, several cluster configurations may yield results equally satisfying from a statistical point of view. Another design choice with technical implications comes up: **are there features that are more appropriate to discriminate against than others?** For instance, if both split yield equivalent results, it is better to split according to age or according to the number of prior convictions? This can be important regarding the **interpretability** of the result. The impact of prior FTA may be more relevant than age when it comes to debating the prediction made by the tool in the light of the defendant's individual context. Therefore, one could define a rule to prompt the algorithm to split preferably along certain features than other. At the aggregate level, nothing would change, but for individual defendants, the choice of the cluster split could have a significant impact on outcomes.

Choosing the “right” split can seem like a quite subjective yet consequential decision. We would thus like to briefly introduce **the Handoff Forest, an extension of the model that would significantly reduce this design subjectivity.** Since there usually are many different trees that look equivalent at the aggregate level, a classic strategy to blur the artificial differences between these models is to use a forest, or a group of trees, combining the different predictions in a final result. It would soften arbitrary threshold effects that an individual located near the border of a cluster would face with a single tree. This approach also **systematically increases accuracy on new data**, and reduces the risk of overfitting. The forest approach would make the model slightly less accountable, as the prediction would no longer be the result of a single combination of features. Nonetheless, it will still be possible to give the weight that each variable or combination of variable had on the outcome. Another addressable but non-trivial problem is to to optimally combine error rates from the different trees, requiring further work to understand how best to train models to maximise various value objectives.

## CONCLUSION

In this report, we first presented a background on state-wide risk assessment tool use, then outlined an analysis of value implications for algorithmic pretrial detention. Informed by a mixed methods research approach, we conclude that while the PSA is a well-intentioned improvement on existing tools, there are multiple areas in which the tool could better support values inherent to human-based decision making: these include addressing inherent disparities, making the model more transparent and less reductionist, as well as exercising caution when building in automatic overrides. Given the very high error rates that pretrial risk assessments entail, we believe that the recommendations must be considered with more critical judgement than they currently are. Significant choices were made during the PSA design to extrapolate variable weights from the training data, which we argue are not sufficiently available to the public, missing sufficient translation of technical language into value tradeoffs that are more easily understood by relevant stakeholders.

We thus offer mitigations that may improve the PSA's implementation, as well as a completely alternative design, the Handoff Tree. This model offers a paradigm shift in that it intelligently and fully delegates decision making to the judge when uncertainty is too high. Considerations about error rates are made an integral part of the prediction reports, which provide nuance and interpretability on what a "high risk" recommendation means, as well as attempting to directly mitigate predictive discrimination. Precision and robustness can also be increased by extending the tree to a forest, though requiring reductions in accountability and interpretability. The design of such a model involves intricate trade-offs, which could lead one to question the value of such an alternative. However, such tussles are inherent to data science, and the way they are addressed is what makes a model accurate and fair.

# APPENDIX

## 1. Sample PSA Court Report

Pretrial Services

Public Safety Assessment - Court Report

Name: [redacted] SF#: [redacted]

DOB: [redacted]

Arrest Date: 9/26/2018 PSA Completion Date: 9/27/2018

New Violent Criminal Activity Flag **No**

| | | | | | | |
|---|---|---|---|---|---|---|
| New Criminal Activity Scale | 1 | 2 | 3 | 4 | 5 | 6 |
| Failure to Appear Scale | 1 | 2 | 3 | 4 | 5 | 6 |

Booked Offense(s):

11351HS/F POSSESS/PURCHASE FOR SALE NARCOTIC/CONTROLLED SUBSTANCE, 11370.1 (A)HS/F POSS CONTROLLED SUBSTANCE WHILE ARMED W/LOADED FIREARM, 29800 (A)(1)PC/F FELON/ADDICT/POSSESS/ETC FIREARM, 30305 (A)(1)PC/F PROHIBITED PERSON OWN/POSSESS/ETC AMMUNITION/ETC

| Risk Factors: | Reponses: |
|---|---|
| 1. Age at Current Arrest | 23 or older |
| 2. Current Violent Offense | No |
| a. Current Violent Offense & 20 Years Old or Younger | No |
| 3. Pending Charge at Time of the Offense | No |
| 4. Prior Misdemeanor Conviction | Yes |
| 5. Prior Felony Conviction | Yes |
| a. Prior Conviction | Yes |
| 6. Prior Violent Conviction | 2 |
| 7. Prior Failure to Appear in Past 2 Years | none |
| 8. Prior Failure to Appear Older than 2 Years | Yes |
| 9. Prior Sentence Incarceration | Yes |

Decision Making Framework Response

| | NCA 1 | NCA 2 | NCA 3 | NCA 4 | NCA 5 | NCA 6 |
|---|---|---|---|---|---|---|
| FTA 1 | OR - NAS | OR - NAS | | | | |
| FTA 2 | OR - NAS | OR - NAS | OR - NAS | OR - MINIMUM | SFPDP - ACM | |
| FTA 3 | | OR - NAS | OR - MINIMUM | Release Not Recommended | SFPDP - ACM | Release Not Recommended |
| FTA 4 | | OR - MINIMUM | SFPDP - ACM | SFPDP - ACM | Release Not Recommended | Release Not Recommended |
| FTA 5 | | SFPDP - ACM | SFPDP - ACM | SFPDP - ACM / Release Not Recommended | Release Not Recommended | Release Not Recommended |
| FTA 6 | | | | Release Not Recommended | Release Not Recommended | Release Not Recommended |

**Release Not Recommended**

Is this Response based on a Step 2 exclusion?

**Yes**

Does this Response include a Step 4 increase?

**No**

## 2. Information Sources Presented in Card-Sorting Activities

| | |
|---|---|
| **PSA Factors** | - Age at current arrest<br>- Current violent offense<br>- Pending charge at the time of the offense<br>- Prior misdemeanor conviction<br>- Prior felony conviction<br>- Prior conviction (misdemeanor or felony)<br>- Prior violent conviction<br>- Prior failure to appear in the past two years<br>- Prior failure to appear older than two years<br>- Prior sentence to incarceration |
| **Assessments** | - Judge's assessment<br>- Defendant's self-assessment<br>- Defendant's family's assessment<br>- Assessment by fellow citizens (jury of peers)<br>- Victim's assessment |
| **Personal Characteristics** | - Gender<br>- Race |
| **Personal Situation** | - Relationship status<br>- Number of dependants<br>- Obligations to care for loved ones<br>- Employed<br>- Number of months employed<br>- Stable housing<br>- Number of months in stable housing<br>- Health conditions<br>- Financial situation |
| **Preparedness for Flight** | - Preparations for traveling outside the country (eg. passport, tickets)<br>- Contacts outside the country |